\documentclass[12pt,draftclsnofoot,onecolumn]{IEEEtran}
\usepackage{fancyhdr}
\usepackage{amsmath,amssymb}
\usepackage{amsfonts}
\usepackage[dvips]{graphicx}
\usepackage{dsfont}
\usepackage{balance}
 \usepackage{algpseudocode}
  \usepackage{algorithm}

\usepackage{enumerate}
\usepackage{graphics}
\usepackage{subfigure}
\usepackage{cite}
\usepackage{mathrsfs}

\newtheorem{theorem}{Theorem}

\newtheorem{corollary}{Corollary}

\newtheorem{definition}{Definition}

\newtheorem{lemma}{Lemma}

\newtheorem{remark}{Remark}

\title{Cloud Compute-and-Forward with Relay Cooperation}
\author{Koralia N. Pappi, \IEEEmembership{Student Member, IEEE,}  Panagiotis~D.~Diamantoulakis, \IEEEmembership{Student Member, IEEE,} Hadi Otrok, \IEEEmembership{Senior Member, IEEE,} and George~K.~Karagiannidis, \IEEEmembership{Fellow, IEEE}
\thanks{K. N. Pappi, P. D. Diamantoulakis, and G. K. Karagiannidis are with the Department of Electrical and Computer Engineering, Aristotle University
of Thessaloniki, 54 124, Thessaloniki, Greece and with the Department of Electrical and Computer
Engineering, Khalifa University, PO Box 127788, Abu Dhabi, UAE (e-mails: \{kpappi,padiamant,geokarag\}@auth.gr)}
\thanks{H. Otrok is with the Department of Electrical and Computer
Engineering, Khalifa University, PO Box 127788, Abu Dhabi, UAE (e-mail: Hadi.Otrok@kustar.ac.ae)}
\thanks{Part of this work has been presented to IEEE Wireless Communications and Networking Conference (WCNC) 2014.}
}
\begin{document}
%

\maketitle

\begin{abstract}
We study a cloud network with $M$ distributed receiving antennas and $L$ users, which transmit their messages towards a centralized decoder (CD), where $M\geq L$. We consider that the cloud network applies the Compute-and-Forward (C\&F) protocol, where $L$ antennas/relays are selected to decode integer equations of the transmitted messages. In this work, we focus on the best relay selection and the optimization of the Physical-Layer Network Coding (PNC) at the relays, aiming at the throughput maximization of the network. Existing literature optimizes PNC with respect to the maximization of the minimum rate among users. The proposed strategy maximizes the sum rate of the users allowing non-symmetric rates, while the optimal solution is explored with the aid of the Pareto frontier. The problem of relay selection is matched to a coalition formation game, where the relays and the CD cooperate in order to maximize their profit. Efficient coalition formation algorithms are proposed, which perform joint relay selection and PNC optimization. Simulation results show that a considerable improvement is achieved compared to existing results, both in terms of the network sum rate and the players' profits.

\end{abstract}

%


\section{Introduction}\label{S:Introduction}

Cloud Radio Access Networks (C-RANs) have recently attracted the interest of both academic research and industry \cite{cran,Flanagan,Lin}. The C-RAN concept is implemented  based on cloud computing and separation of radio antennas and base stations (BSs), with the latter being replaced by one or more cloud Base Stations (CBSs). Several operations traditionally performed at the BSs, such as network management, handover management, cooperation enforcement etc., are now moved to the CBS. Therefore, this approach decreases capital and operational costs, since it better takes advantage of the processing power of the CBS, which comprises distributed processing resources (e.g. data centers in different locations). Thus, the traditional communication network is replaced by the shared communication and processing infrastructure of the cloud, which is handled by different operators in a distributed manner. However, C-RAN architecture simultaneously arises many design and implementation challenges caused by the limitations of wireless links \cite{Vassilaras}.

C-RAN architecture involves wireless cloud networks comprised by multiple nodes, which transmit their messages simultaneously. Various techniques have been proposed for forwarding messages in these networks, such as Compress-and-Forward \cite{Cover}, Quantize-reMap-and-Forward (QMF) \cite{Avestimehr} and Noisy Network Coding \cite{Lim}. Most of these techniques are complex, while the additive noise builds up with each retransmission, since decoding is not performed at intermediate nodes. A promising alternative is Compute-and-Forward (C\&F) relaying, first introduced in the pioneering work of B. Nazer and M. Gastpar \cite{Nazer} for general relay networks.  This relaying technique exploits interference in multiuser wireless networks, by decoding integer equations of the transmitted coded messages by multiple users, using nested lattice codes \cite{Erez}. The decoded equations are then forwarded by the relays towards a Centralized Decoder (CD) which, having enough independent equations, decodes the transmitted messages from all users.
The choice of integer equation coefficients at each C\&F relay is performed with the use of Physical Layer Network Coding (PNC)\cite{Zhang}, a special case of Network Coding (NC) \cite{Ahlswede,Medard}, which makes use of the additive nature of the wireless medium to combine different source messages, simultaneously arriving at the destination. The C\&F relaying protocol is an attractive choice for cloud networks that aim at reducing the complexity at the backbone by decentralized processing at the relaying nodes. Each node performs lattice decoding, a rather simple decoding technique, which does not require a complex node architecture (e.g. a quantizer, a Maximum-Likelihood decoder etc.).  However, when multiple relays forward equations to the CD, a joint optimization of the PNC must be performed.
\pubidadjcol

In future cloud networks with shared infrastructure \cite{WangWang}, the most challenging case is when the antennas and the cloud backhaul of the network belong to different operators, which want to maximize their profit. Game theory provides a formal analytical framework with a set of mathematical tools to study the complex interactions among rational players. Game theory is employed in numerous works which investigate network selection in heterogeneous networks \cite{Trestian,Fux} and competition or cooperation among operators \cite{Duan,Antoniou,Xing}. Furthermore, game theory has emerged as a tool for communication network analysis and specifically network coding \cite{Saad,Han,ZhangLi}, providing several approaches for the strategy of the network nodes-players, individual or cooperative. Specifically in a C\&F network, in the lack of a single BS which can accommodate all users, multiple relays serve the users, which need to cooperate in order to increase the profit of their operators. Thus, the behavior and interactions of self-organizing C\&F relays can be analyzed using game theoretical tools, especially coalitional games.

\vspace{-0.2in}
\subsection{Motivation}\label{SS:Motivation}
This work is motivated by the complex problem of PNC optimization at the relays, in the uplink of wireless cloud network, employing C\&F. When a C\&F relay chooses the equation coefficients independently, a criterion which defines its strategy can be the maximization of the achievable computation rate region \cite{Nazer}. This maximization has no analytical solution, but various works have proposed algorithmic solutions to this problem, using Lenstra-Lenstra-Lov\'{a}sz (LLL) lattice reduction \cite{Feng2}, geometric programming \cite{Sussi} and a modified Fincke-Pohst method \cite{Wei2}, among others. However, when various C\&F relays choose their equations independently, a set of linearly independent equations is not guaranteed at the CD's side, the sum rate is not maximized and the optimization of individual computation rates does not lead to the maximization of each operator's profit.

In \cite{Wei2}, the authors propose an optimization of the equation coefficient vectors, so that the relays provide the CD with a set of independent equations. The optimization criterion is the maximization of the minimum transmission rate among the sources. Similar strategy was followed in \cite{Mejri} for Complete-C\&F and Incomplete-C\&F. However, although these works optimize the PNC without resorting to an exhaustive search for the first time, both approaches do not address the maximization of the total throughput, i.e. the sum rate. This is a crucial metric of the network performance, since the users usually pay according to the data volume they transmit or receive.

The optimization of the sum rate is partially dealt with in \cite{Caire}, \cite{Caire2} and \cite{Soussi2}. In \cite{Caire,Caire2}, relay and user selection for a CBS employing C\&F and its low complexity version, named Quantized C\&F \cite{Hong}, are investigated. For the uplink, the relays choose their equation coefficients independently and after a decomposition of the network into subnetworks, a greedy algorithm within each subnetwork for the best selection of relays, is performed. However, although the solution in \cite{Caire} and \cite{Caire2} is ingenious, it is in general suboptimal, since it considers only symmetric transmission rates within each subnetwork, while the relays always choose one specific coefficient vector each, beforehand. Therefore, no cooperation or tradeoff between relays for the rates' optimization can be performed. The authors in \cite{Soussi2} also concentrate on the symmetric rate case, not allowing non-symmetric transmission rates, a choice which can lead to an overall increase of the total throughput. The need for cooperation calls for the formulation of proper incentives and corresponding rules that will model the cooperation between relays, which can be offered by a game theoretic framework.
\vspace{-0.2in}
\subsection{Contribution}\label{SS:Contribution}
In this work, the uplink of a C\&F Cloud Base Station (CBS) is considered, employing surplus relays with respect to the number of the sources. A practical scenario in which neither the relays (distributed antennas) not the cloud CD belong to the same operator is considered, which is crucial for future distributed heterogeneous cloud-based networks. Moreover, operator profits are introduced, which are based on the provided services to the users. Thus, in this work we establish the incentive for cooperation among the relays and the cloud CD by formulating a coalition game, leading to a joint optimization of PNC.

Notice that previous studies aim at the maximization of the minimum transmission
rate\cite{Wei2,Mejri} or the symmetric transmission rate. In this paper, we consider the sum rate allowing non-symmetric rates at the sources, which is the most important metric of the network performance. Furthermore, since the maximization of the minimum transmission rate and the maximization of the sum rate are two objectives which may be conflicting and non-commensurable, we use the Pareto frontier to define the dominating solutions and to show the tradeoff between the maximization of the two rates.

In order to investigate the cloud's incentive to reach the optimal solution in terms of sum rate, coalitional game theory is applied, aiming at the maximization of the cloud's revenue. The impact of relay selection in a C\&F network is also investigated. Extensive simulations over fading channels have been conducted to evaluate the performance of the proposed coalition formation algorithms.

The contribution of this work is summarized in the following:
\begin{itemize}
\item The optimal relay selection and PNC, which maximize the sum rate of the C\&F network under constraints for the minimum rate, is studied with the use of Pareto frontier.
\item  The relay selection is matched to a game-theoretical coalition formulation problem which aims at the maximization of the cloud revenue, while the profits of the relays' and the CD's operators are introduced.
\item Efficient coalition formation algorithms are proposed, which converge to a solution belonging to the core. As illustrated by the results, the proposed method leads to great sum rate gains compared to previous works, with simultaneous optimization of the cloud revenue. It is remarkable that the algorithmic optimization may be more beneficial compared to just adding more relays to the network.
\end{itemize}
\vspace{-0.2in}
\subsection{Organization}
The rest of the paper is organized as follows: In Section
\ref{S:System_model}, the system model, the achievable rates of
a C\&F network and the problem statement are presented. In Section \ref{S:Network_Management}, the joint network coding optimization is studied with the use of Pareto frontier.
The game model and the proposed coalition formation algorithms are presented in Sections \ref{S:Game} and
\ref{S:Algorithms} respectively. Specific
examples and simulation results are discussed in Section \ref{S:Examples} and Section \ref{S:Simulation_results} respectively, while the
conclusions are given in Section \ref{S:Conclusions}.
\vspace{-0.2in}
\section{System Model}\label{S:System_model}
\subsection{Cloud Compute-and-Forward Relaying}\label{SS:Compute_and_forward_network}
\begin{figure}[t!]
\centering\includegraphics[keepaspectratio,width=0.4\columnwidth]{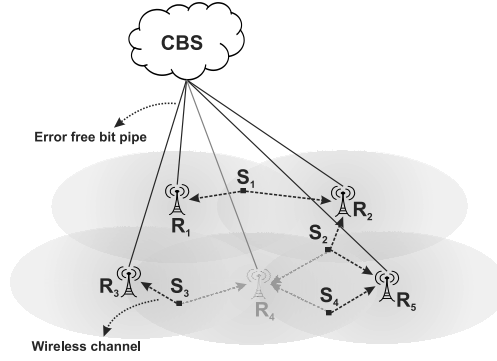}
\caption{Cloud Compute-and-Forward network with relay selection.}
\label{Fig:CF_network}
\end{figure}

We consider the uplink of a real-valued\footnote{The results of this work can be directly extended to complex-valued systems (see also \cite{Nazer}).} cloud C\&F network, which consists of two units, i) the baseband unit (BBU), which corresponds to the CBS and contains the CD of the C\&F scheme, and ii) the remote radio head (RRH), which corresponds to the available relays or antennas \cite{cran}. The CBS receives messages from $L$ sources, denoted by $S_l$, $l=1,2,\ldots,L$, which transmit messages encoded using a lattice code \cite{Erez}, while the RRH consists of $M$ relays, denoted by $R_m$, $m=1,2,\ldots,M$, where $M\geq L$. As shown in Fig.\ref{Fig:CF_network}, the relays are connected to the CBS via error free bit pipes as in \cite{Nazer} (e.g. optical fibers with fixed rate $R_0$ which is considered to be much higher than the rates achieved over the wireless medium).

The channel matrix between the sources and the relays $\mathbf{H}$ is an $L\times M$ matrix, given by
\begin{equation}\label{channel_matrix}
\mathbf{H}=[\mathbf{h}_1, \mathbf{h}_2, \ldots, \mathbf{h}_M],
\end{equation}
where $\mathbf{h}_m$ is the channel coefficient vector of length $L$ as seen by the relay $R_m$, that is
\begin{equation}
\mathbf{h}_m=[h_{m,1}, h_{m,2}, \ldots, h_{m,L}]^T,
\end{equation}
where $h_{m,l}$ is the real channel coefficient between the source $S_l$ and the relay $R_m$, and $[\cdot]^T$ denotes a matrix or vector transpose. Note that, in the case of complex Gaussian circularly symmetric fading channels, the real channel coefficients are produced by the projection of the complex channel on the in-phase axis, and thus, they follow a Gaussian distribution, which is considered here as $\mathcal{N}(0,1)$. Each relay has perfect CSI of the channels between the sources and the specific relay, but no CSI is considered at the transmitters \cite{Nazer}.

In \cite{Nazer}, a C\&F relay performs lattice decoding, thus, each is comprised by two units, the radio function unit (RFU) which filters and possibly converts the frequency of the received signal, and the signal processing unit (SPU) which performs the baseband processing of the signal.
At the RFU, each relay receives a signal
\begin{equation}\label{received_signal}
\mathbf{y}_m=\sum\limits_{l=1}^Lh_{m,l}\mathbf{x}_l+\mathbf{z}_m,
\end{equation}
where $\mathbf{x}_l\in\mathbb{R}^N$ is the signal vector transmitted by the source $S_l$, $\mathbf{y}_m\in\mathbb{R}^N$ is the received vector by the relay $R_m$, $N$ is the dimension of the nested lattice code \cite{Erez} used in the C\&F scheme and $\mathbf{z}_m\in\mathbb{R}^N$ is the noise vector which follows a normal distribution $\mathcal{N}\left(\mathbf{0}, \mathbf{I}_{N\times N}\right)$, with $\mathbf{0}$ being the zero vector of length $N$ and $\mathbf{I}_{N\times N}$ being the $N\times N$ identity matrix. Each transmitted signal is subject to a power constraint $P$, i.e. $\frac{1}{N}\|\mathbf{x}_l\|^2\leq P$, where $\|\cdot\|$ denotes the vector norm.

The SPUs of the relays decode integer equations of the transmitted vectors, with equation coefficients given by the vectors
\begin{equation}\label{equation_vector}
\mathbf{a}_m=[a_{m,1}, a_{m,2}, \ldots, a_{m,L}]^T,\quad\mathbf{a}_m\in\mathbb{Z}^L.
\end{equation}
Each relay, thus, decodes a codeword
\begin{equation}\label{codeword}
\hat{\mathbf{x}}_m=\left[\sum\limits_{l=1}^La_{m,l}\mathbf{x}_l\right]\mathrm{mod}\Lambda,
\end{equation}
where $[\cdot]\mathrm{mod}\Lambda$ is the modulo lattice operation on the coarse lattice of the nested lattice code in use \cite{Nazer,Erez}.

The CD can decode all messages if it is provided with $L$ independent equations. The CD inverts the $L\times L$ network coding matrix $\mathbf{A}$, whose columns are the equation coefficient vectors $\mathbf{a}_m$ of the $L$ selected relays, so the network coding matrix $\mathbf{A}$ must be full rank:
\begin{equation}\label{matrix_A}
\mathbf{A}=[\mathbf{a}_{m1}, \mathbf{a}_{m2},\ldots\mathbf{a}_{mL}],
\end{equation}
where $m1,m2,\ldots,mL$ are the indices of the $L$ selected relays. Note that the entries of $\mathbf{A}$ are integer numbers. We consider $\mathbf{a}_m=\mathbf{0}$ for the relays which are not selected.

\begin{figure}[t!]
\centering\includegraphics[keepaspectratio,width=0.4\columnwidth]{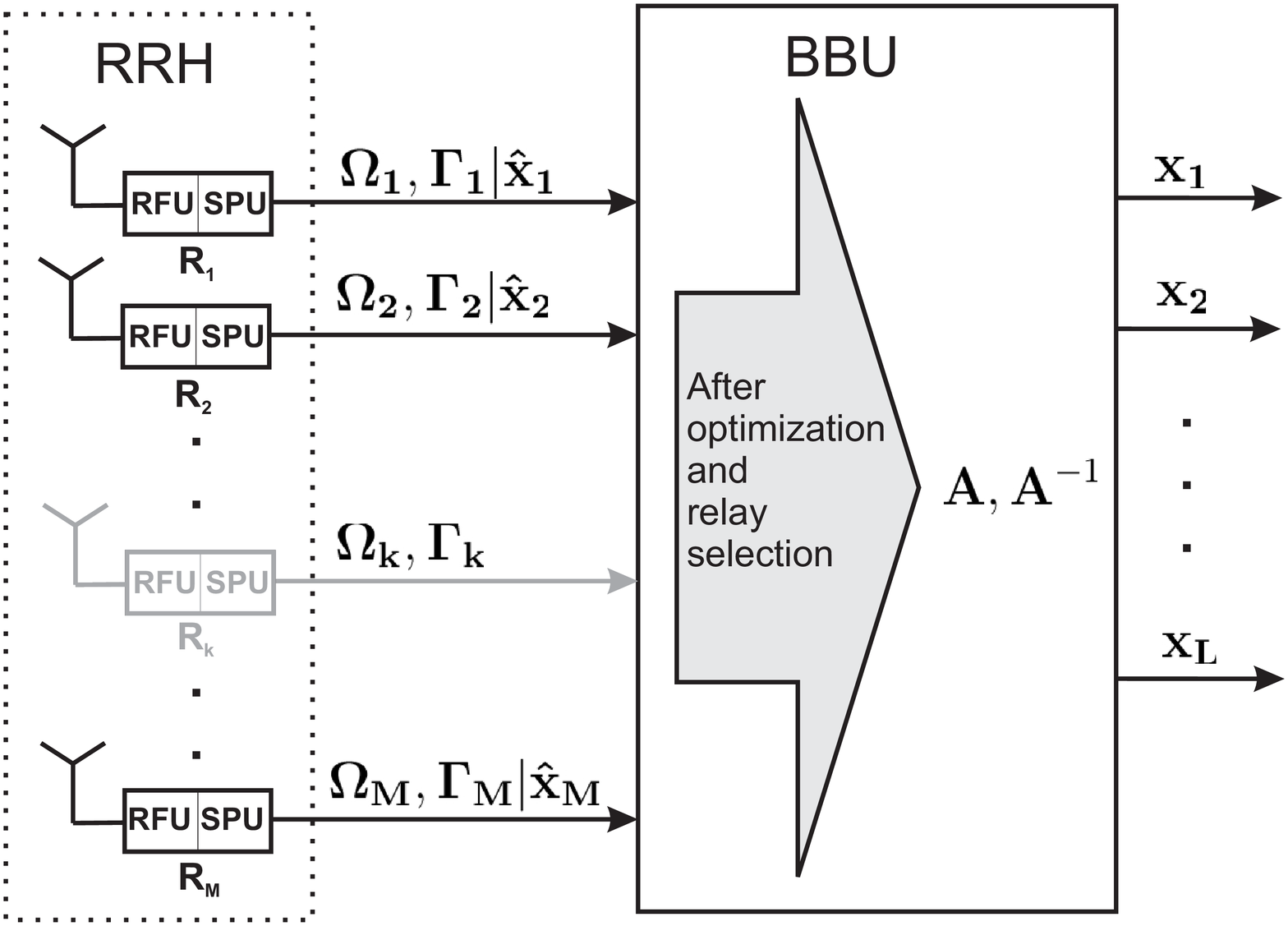}
\caption{Partial centralization architecture with relay selection.}
\label{Fig:partial}
\end{figure}

According to the different function splitting between the BBU and the RRH, there are two cloud architectures, named partial centralization and full centralization \cite{cran}. We adopt partial centralization, as depicted in Fig. \ref{Fig:partial}, where the RRH integrates both the radio function and the baseband signal processing performed by the SPUs. Thus, the relays can communicate their candidate equation coefficient vectors and achievable rates during the optimization process through the cloud. After relay selection, only the selected relays perform equation decoding and forward the estimated codewords $\hat{\mathbf{x}}_m$ in (\ref{codeword}) towards the CD, while the rest of the relays remain silent or turn off, as indicated by gray color in Fig. \ref{Fig:partial}. The CD corresponds to the BBU and performs the inversion of the network coding matrix $\mathbf{A}$ and the decoding of the messages transmitted by the sources.
While the relays and the CD are cooperating nodes, there is no need for wireless overhead, since all information regarding the choice of the equation coefficient vectors $\mathbf{a}_m$ is circulated via the error free bit pipes which connect the RRH to the BBU. More specifically, each relay needs to send each own set of candidate vectors and corresponding achievable rates towards the CD, only when the channel coefficients change.

%
\vspace{-0.2in}
\subsection{Compute-and-Forward Rate Region}\label{SS:Compute_and_forward_rates}
The choice of the equation coefficient vector determines the rate that a relay can achieve, while decoding equations of messages. It was proven in \cite{Nazer} that the achievable computation rate region of a relay $R_m$ for a given choice of $\mathbf{a}_m$ and given channel coefficients is
\begin{equation}\label{computation_rate}
\mathcal{R}^r_m=\frac{1}{2}\log_2^+\left(\left(\|\mathbf{a}_m\|^2-\frac{P\left(\mathbf{h}_m^T\mathbf{a}_m\right)^2}{1+P\|\mathbf{h}_m\|^2}\right)^{-1}\right),
\end{equation}
where $\log_2^+(\cdot)=\max\left(\log_2(\cdot),0\right)$ and $(\cdot)^r$ denotes the computation rate of a relay. The maximization of the rate in (\ref{computation_rate}) with respect to the equation coefficient vector $\mathbf{a}_m$ is achieved when \cite{Feng2}
\begin{equation}\label{optimal_eq_coef_vector}
\mathbf{a}_m=\mathrm{arg}\min_{\mathbf{a}\in\mathbb{Z}^L,\mathbf{a}\neq \boldsymbol{0}}\mathbf{a}^T\mathbf{G}\mathbf{a},
\end{equation}
where $\mathbf{G}=\mathbf{I}_{L\times L}-\frac{P\mathbf{h}_m\mathbf{h}_m^T}{1+P\|\mathbf{h}_m\|^2}$.

Now, each transmitter encodes a message $\mathbf{w}_l\in\mathbb{F}_p^{k_l}$, where $\mathbb{F}_p$ is a $p$-sized field with $p$ being a prime, onto a lattice codeword $\mathbf{x}_l\in\mathbb{R}^N$. Thus, the rate of each transmitter is $\mathcal{R}^s_l=\frac{k_l}{N}\log_2p$, where $(\cdot)^s$ denotes a transmission rate from a source. For a given achievable rate of each source, suitable choices for the values of $k_l,N$ and $p$, must be found, as discussed in \cite{Nazer}.
The transmission rate of a given source is bounded by the minimum
computation rate among only those relays which decode an equation
containing the codeword transmitted by this source. That is
\cite{Nazer},
\begin{equation}\label{transmit_rate_2}
\mathcal{R}^s_l\leq\min_{a_{m,l}\neq 0}\mathcal{R}^r_m.
\end{equation}
\vspace{-0.2in}
\subsection{Problem Statement}\label{SS:Problem_statement}

Previous works \cite{Wei2,Mejri} have performed the optimization of the PNC in C\&F networks, based on the maximization of the minimum computation rate, which implies a symmetric transmission rate scenario. This optimization ensures that no user will transmit with very low rate. However, this optimization does not guarantee that the total throughput of the network is also maximized, especially in the case of non-symmetric transmission rates, since according to (\ref{transmit_rate_2}), the $l$-th user's transmission rate is defined only by those equation coefficient vectors with non-zero elements at the $l$-th position. The above lead to the following two observations:

\begin{remark}\label{Remark:diff_transm_rates}
In a network with symmetric rates, the common transmission rate is the minimum computation rate. When non-symmetric transmission rates are allowed, a vector $\mathbf{a}\in\mathbf{A}$, which achieves the minimum computation rate, defines
only those transmission rates corresponding to its non-zero
elements, while the rest of the rates are defined by other vectors in $\mathbf{A}$ which achieve higher rates.
\end{remark}
\begin{remark}\label{Remark:sum_vs_min}
When non-symmetric rates are considered, the maximization of the sum rate is not equivalent to the maximization of the minimum transmission rate. Choosing a vector
$\mathbf{a}$ which contains many zero elements and defines the minimum rate may be favorable compared to another with less zero elements and higher computation rate. This shows that there are cases when decreasing the minimum rate may increase the sum rate.
\end{remark}

In a network where each operator of an antenna (relay) or of the CD is rewarded according to the throughput that it serves, the maximization of the total throughput is of utmost importance. Furthermore, each user does not have the same needs for high transmission rates. For example, a user uploading a sort message (e.g. a tweet) and a user uploading a video, do not require the same upload rate from the service provider. All of the above lead to the need of maximizing the sum rate of the network, under specific constraints on the minimum transmission rate that each user accepts. Thus, the tradeoff between minimum transmission rate and sum rate is considered, while the optimal solution is examined in Section \ref{S:Network_Management}.

The optimal solution in terms of sum rate cannot be acquired, without the collaboration of the relays. If each relay individually transmits its best equation in terms of computation rate, the following may occur: i) The availability of a set of $L$ independent equations is not guaranteed. If there are not enough independent equations, then the system is in outage. ii) If each relay chooses only its best equation in terms of computation rate, a) it may not contribute to the final coding matrix $\mathbf{A}$, in which case it does not serve any throughput, and thus it gains no profit, b) if it contributes to the coding matrix, there may be another solution, in which it could contribute with another equation, achieving higher total throughput. In the latter case, its profit would be higher.

Similar observations are made for the CD. If it randomly chooses $L$ equations, its profit is not necessarily maximized. Thus, the problem naturally leads to the need of cooperation among relays and CD, with the aim to achieve the optimal solution in terms of sum rate. Therefore, the problem is matched to a coalition formation game, where each player seeks to form a coalition which achieves the maximum sum rate and the maximum profit for the relays and the CD. The game is formulated in Section \ref{S:Game}.

\section{Network Management}\label{S:Network_Management}

\subsection{Physical Layer Network Coding Optimization Criterion}\label{SS:Criteria}

In practical networks, the requirements of each user in terms of rate
are different (e.g. they use different kind of services etc.). Furthermore, the prices paid by the users for the services of the operators are proportional to the rates they achieve. Thus, an
optimal strategy for such a network is the maximization of the sum rate under the fulfilment of minimum requirements for each
user. This criterion can be
formulated as
\begin{equation}\label{criterion3}
\begin{array}{ll}
\textbf{find}&\mathbf{A}=\mathrm{arg}\max\limits_{|\mathbf{A}|\neq 0}\left(\sum\limits_{l=1}^L\mathcal{R}^s_l\right)\\
\textbf{subject
to}&\mathcal{R}^s_l\geq V_l,\,\,\,l=1,2,\ldots,L,
\end{array}
\end{equation}
where $V_l$ is minimum rate requirement of the $l$-th source. In case when $V_l=V$, $\forall l$, then the optimization criterion corresponds to the maximization of the sum rate, under a constraint for the minimum rate among users.

\subsection{The Pareto Frontier}\label{SS:Pareto_frontier}
In this section, we investigate the case when the minimum rate requirement is the same for all users, i.e. $V_l=V$, $\forall l$. Often, the maximization of the sum rate and the minimum transmission rate are two conflicting strategies, as explained in Remark \ref{Remark:sum_vs_min}. To better investigate and face this conflict, we utilize the concept of Pareto frontier, which is defined formally as follows.

\begin{definition} \textbf{Pareto frontier:}
The Pareto frontier is the set of matrices that are not strictly
dominated by any other matrix. A matrix $\mathbf{A}'$ is said to
Pareto dominate another solution matrix $\mathbf{A}$
$\left(\mathbf{A}'\succ\mathbf{A}\right)$, if
\begin{equation}
\left(\min\limits_{m:\mathbf{a}_m\in\mathbf{A}'}\mathcal{R}_m^r\geq\min\limits_{m:\mathbf{a}_m\in\mathbf{A}}\mathcal{R}_m^r\right)\cap\left(\sum\limits_{\begin{subarray}{c}l=1\\\mathbf{a}_m\in\mathbf{A}'\end{subarray}}^L\mathcal{R}_l^s>\sum\limits_{\begin{subarray}{c}l=1\\\mathbf{a}_m\in\mathbf{A}\end{subarray}}^L\mathcal{R}_l^s\right),
\end{equation}
or
\begin{equation}
\left(\min\limits_{m:\mathbf{a}_m\in\mathbf{A}'}\mathcal{R}_m^r>\min\limits_{m:\mathbf{a}_m\in\mathbf{A}}\mathcal{R}_m^r\right)\cap\left(\sum\limits_{\begin{subarray}{c}l=1\\\mathbf{a}_m\in\mathbf{A}'\end{subarray}}^L\mathcal{R}_l^s\geq\sum\limits_{\begin{subarray}{c}l=1\\\mathbf{a}_m\in\mathbf{A}\end{subarray}}^L\mathcal{R}_l^s\right).
\end{equation}
\end{definition}

Formally, this defines a partial order on all possible matrices
$\mathbf{A}$, and the Pareto Frontier is the set of maximal
elements with respect to this order. Note that there may be multiple matrices $\mathbf{A}$ which achieve the same sum rate under specific minimum rate requirements. In that case, these matrices are equivalent and the selection of any of these matrices does not affect the performance of the network and the players' payoffs, as they will be defined in the next section.

When there is a matrix $\mathbf{A}$ which Pareto dominates all the
other possible matrices, then the Pareto Frontier consists of only
one element, the Pareto optimal point. When such a point is
available, then this matrix maximizes simultaneously both the
minimum rate and the sum rate of the network. However, when there is no Pareto optimal point, then it is obvious that the two optimization criteria lead to a different network coding solution.

In the case of different $V_l$ for each source, the matrices $\mathbf{A}$ which do not achieve all minimum requirements set by the users, are excluded by the set of solutions beforehand, as it will be described in detail in Section \ref{S:Algorithms}. Then, the optimization is again performed considering the tradeoff between minimum and sum rate.

The points on the Pareto frontier achieve different sum rate. The solution of the optimization problem in (\ref{criterion3}) is the point on the Pareto frontier which achieves the maximum sum rate, while at the same time, it holds that $\min\limits_{m:\mathbf{a}_m\in\mathbf{A}}\mathcal{R}_m^r\geq V$.

\section{PNC as a Coalitional Game For Cooperation Among Relays and Cloud CD}\label{S:Game}
In this section, a very challenging scenario is examined, in which neither the relays (distributed antennas) not the cloud CD belong to the same operator. The practical meaning behind this is that the communication operators will make profit for providing
services,while the price paid by the users will compensate the use of the relays and the use of the cloud CD, yielding the corresponding profit to the providers of the respective infrastructure. This separation is a fundamental step toward the distributed heterogeneous cloud-based networks. The objective of this section is to investigate which is the incentive for cooperation among the relays and the cloud CD, and the formation of coalitions.
\subsection{Coalition Formation Game: Basics}\label{SS:Game_basics}

For the purpose of deriving an algorithm that allows the relays
to form coalitions in a distributed manner, we
use notions from cooperative game theory. In this regard, we formulate the
cooperative relay selection problem of the previous section
as a coalitional game in characteristic form with transferable
utility (TU).
Below we define the game and its essential elements \cite{Han}.
\begin{definition}\label{defining} \textbf{Game:}
A coalitional game with TU
is defined by a pair $(\mathcal{N}, v)$ where $\mathcal{N}=\mathcal{M}\cup\mathcal{C}$ is the set of players which includes the set $\mathcal{M}$ of available relays and the cloud CD, denoted by $\mathcal{C}$. The function
$v$ is defined over the real line such that for every coalition
$\mathcal{S} \subseteq \mathcal{N}$, $v(\mathcal{S})$ is a real number describing the amount of utility that coalition $\mathcal{S}$ receives and which can be distributed in any arbitrary manner among the members of the coalition. The
\textit{information} available at each decision point includes the set of candidate vectors and corresponding computation rates. The
\textit{actions} available to the players at each decision point
are whether they will enter or leave a coalition, and which
candidate equation coefficient vector the relays will choose.
\end{definition}
%

Both the relays and the cloud CD are paid by the sources proportionally to the offered QoS. When the objective of the sources is the maximization of their rate, then each user pays according to the achieved transmission rate, so the total payment received by the cloud depends on the sum rate. In this case, the revenue, i.e. the total payment that a coalition receives per channel use, is given by
\begin{equation}\label{revenue2}
v(\mathcal{S})=\begin{cases}  Z\sum\limits_{l=1}^L\mathcal{R}^s_l, & |\mathcal{S}|\geq L+1, \mathcal{C} \in \mathcal{S}, \mathcal{R}^s_l\geq V_l,\,\forall l\\
                     0, & \textit{otherwise}
                     \end{cases}
\end{equation}
where $Z$ is the payment of the users concerning the use of relays and the use of the cloud CD, per exchanged bit. By (\ref{revenue2}), it is evident that the revenue is non-zero, when there are at least $L$ available equations, the CD is in the coalition in order to detect the original messages, while at the same time the QoS constraints of the users are met. The last constraint in (\ref{revenue2}) implies that the users will not pay if their transmission rate of the network is too low. When the constraints are not met, then neither the relays nor the cloud CD will be rewarded, and thus they are forced to form a new partition which satisfies these constraints. In the case of the existence of a user whose QoS requirements cannot be met in any way, this user is discarded and the rest of the users are served by the cloud (a new number $L$ of users is selected and the game is repeated). Finally, only the coalition that includes the cloud CD will be rewarded, i.e. the coalition of relays which cooperate with the CD towards the decoding of the transmitted messages.

Since (\ref{revenue2}) represents a revenue gained by the coalition, i.e., a certain
amount of money, it can thus be divided in any arbitrary manner
between the members of $\mathcal{S}$, which implies that we have a game
with transferable utility. Although a number of fairness criteria
(e.g., egalitarian fair, Shapley value, nucleolus, etc.) exist for the
division of payoffs, we consider that the payment is firstly divided among the cloud CD and the relays, while among the relays we adopt the equal fair allocation rule. Thus, when the TU is non-zero, the payoff of any player $i \in \mathcal{S}$, denoted by $\phi_i(\mathcal{S})$ is
\begin{equation}\label{revenue_players}
\phi_i(\mathcal{S})=\begin{cases} \frac{b}{(|\mathcal{S}|-1)}v(\mathcal{S}), & i\neq \mathcal{C}\\
(1-b)\,v(\mathcal{S}), & i=\mathcal{C}
\end{cases}
\end{equation}
where $0<b$. Thus, $b$ is the portion of the TU concerning the revenue of the relays and $(1-b)$ is the portion concerning the revenue of the cloud CD. The payment concerning the relays is equally divided among the relays (i.e., $\frac{b}{|\mathcal{S}|-1}$, since $|\mathcal{S}|-1$ is the number of relays in the coalition), because each relay contributes in the same way for the decoding of each user's message.

Note that the information between players is exchanged
through the CD and via error free bit pipes, thus there is no
need for wireless interaction between the relays. In fact, the interaction between players at each decision point may be virtual. In this case, the players forward their sets of candidate equation vectors and achievable rates only once to the BBU, where the game is played in a centralized manner, as depicted in Fig. \ref{Fig:partial}.


\subsection{Preference Relations and Core Partition}\label{SS:Game_core}

We first cite the following definitions \cite{Banerjee}, for sake of completeness of our analysis.

\begin{definition}\label{preference_relation_def}\textbf{Preference relation:}
A preference relation, denoted by $\succeq_i$ is a reflexive, complete and transitive binary relation on $\mathcal{S}_i(\mathcal{N})=\{\mathcal{S}\in 2^{\mathcal{N}}:i\in\mathcal{S}\}$. Strict preference relation and the indifference relation are denoted by $\succ_i$ and $\sim_i$ respectively $(\mathcal{S}\succ_i \mathcal{T}\Longleftrightarrow [\mathcal{S}\succeq_i \mathcal{T} \text{ and } \mathcal{T}\not\succeq_i \mathcal{S}] \text{ and } \mathcal{S}\sim_i \mathcal{T}\Longleftrightarrow [\mathcal{S}\succeq_i \mathcal{T} \text{ and } \mathcal{T}\succeq_i \mathcal{S}])$.
\end{definition}

\begin{definition}\label{partitions_def} \textbf{Partition:}
A coalition structure $\pi=\{\mathcal{S}_1,\mathcal{S}_2,\ldots,\mathcal{S}_K\}$, ($K\leq|\mathcal{N}|$ is a positive integer) is a partition of $\mathcal{N}$. That is, $\mathcal{S}_k\neq\emptyset$ for any $k\in\{1,2,\ldots,K\}$, $\bigcup_{k=1}^K\mathcal{S}_k=\mathcal{N}$, and $\mathcal{S}_l\cap\mathcal{S}_k=\emptyset$ for any $k,l\in\{1,2,\ldots,K\}$ with $k\neq l$. For any coalition structure $\pi$ and any player $i$ let $\pi(i)=\{\mathcal{S}\in\pi:i\in\mathcal{S}\}$ be the set of their partners. The collection of all coalition structures in $\mathcal{N}$ is denoted by $\Pi(\mathcal{N})$.
\end{definition}

\begin{definition}\label{core_def} \textbf{Core partition:}
A coalition $\mathcal{T}\in2^{\mathcal{N}}\backslash\{\emptyset\}$ is a profitable coalitional deviation from $\pi\in\Pi(\mathcal{N})$ iff $\mathcal{T}\succeq_i\pi(i)$ for any $i\in \mathcal{T}$. A core partition is a partition $\pi^*$ that is immune to any coalitional deviation.
\end{definition}

\begin{theorem}\label{stable_game}
The game $(\mathcal{N}, v)$, as described in section \ref{SS:Game_basics}, has a non-empty core.
\end{theorem}
\begin{IEEEproof}
The proof is given in Appendix A.
\end{IEEEproof}

Following, we also prove that a specific partition of the players is a core partition.

\begin{theorem}\label{core_partition}
Let $\mathcal{S}^*\subseteq\mathcal{N}$ be a coalition which maximizes the sum rate, contains exactly $L$ relays and the CD, that is, $\mathcal{S}^*=\{\mathcal{S}^*\ni\mathcal{C},|\mathcal{S}^*|=L+1, v(\mathcal{S}^*)=Z\max\left(\sum_{l=1}^L\mathcal{R}_l^s\right)\}$. The partition $\pi^*=\{\mathcal{S}^*, \mathcal{S}_1,\ldots\mathcal{S}_{M-L}:|\mathcal{S}_i|=1, i=1,\ldots,M-L\}$ is a core partition, that is, the partition containing the coalition $\mathcal{S}^*$ and all the rest of the players as individuals (if any), is immune to any profitable coalitional deviation.
\end{theorem}
\begin{IEEEproof}
The proof is given in Appendix B.
\end{IEEEproof}

We note here that the rules of \textit{split and merge} or \textit{switch} do not guarantee that the game will reach a core partition, since they pose strong assumptions on the limits of mobility that the players have \cite{Saad}. Moreover, the aforementioned rules do not dictate which is the best choice of equation coefficient vector for each relay, even if a coalition is formed. Thus, in the following section we present a coalition formation algorithm where the players care about the size of their coalition, as dictated by Theorem \ref{core_partition}, and also choose the equation coefficient vectors that maximize the sum rate within the coalition. Considering an ordering of the candidate equation coefficient vectors of all players according to the achievable computation rates as was done in \cite{Wei2}, the proposed algorithm forms coalitions of specific size. The players join the coalition sequentially, as in \cite{Bogomolnaia}, based on the order of the candidate equation coefficient vectors.

\vspace{-0.15in}

\section{Coalition Formation Algorithm}\label{S:Algorithms}
In this section, we present an algorithmic strategy which aims to find the optimal partition of the players and the optimal coding vectors within the coalition which contains the CD, as described in Section \ref{S:Game}. We first describe the formation of the candidate equation coefficient vectors which are available for each relay, and then we present the coalition formation strategy of the players.
\vspace{-0.15in}
\subsection{Formation of the candidate equation coefficient vector sets}\label{SS:candidate_sets}

The problem of finding the best equation coefficient vector given in  (\ref{optimal_eq_coef_vector}) can be mapped to finding the shortest vector on a lattice with generator matrix $\mathbf{G}$, as in \cite{Feng2}. Thus, various algorithms are used, which originate in applications of lattice theory, such as the Fincke-Pohst method \cite{Fincke}, a modified version of which was used in \cite{Wei2}, or LLL lattice basis reduction with Schnorr - Euchner enumeration which was used in \cite{Caire2}. These algorithms can be also used to find a set of candidate coefficient vectors, denoted by $\Omega_m$, which are the best in terms of achievable computation rate of the $m$-th relay.

In this section, we define the appropriate length of these sets based on various criteria, while we further eliminate some of the candidate vectors found by the aforementioned algorithms in the literature, which can be a priori excluded, since they do not contribute to the achievable rate or the full rank property of $\mathbf{A}$. To this end, the following lemma is first presented.
\begin{lemma}\label{Lemma1}
If $\mathbf{a}_1$ is a candidate equation coefficient vector corresponding to a channel coefficient vector $\mathbf{h}$, and
$\mathbf{a}_2=\lambda\mathbf{a}_1$, where
$\lambda\in\mathbb{Z},\,|\lambda|>1$, then the computation rate
achieved using $\mathbf{a}_1$ is greater than the one achieved
using $\mathbf{a}_2$.
\end{lemma}
\begin{IEEEproof}
The computation rate using $\mathbf{a}_2$ is calculated as
\begin{equation}
\begin{split}
\mathcal{R}^r\left(\mathbf{a}_2\right)&=\frac{1}{2}\log_2^+\left(\left(\|\mathbf{a}_2\|^2-\frac{P\left(\mathbf{h}^T\mathbf{a}_2\right)^2}{1+P\|\mathbf{h}\|^2}\right)^{-1}\right)\\&=\frac{1}{2}\log_2^+\left(\lambda^{-2}\left(\|\mathbf{a}_1\|^2-\frac{P\left(\mathbf{h}^T\mathbf{a}_1\right)^2}{1+P\|\mathbf{h}\|^2}\right)^{-1}\right)
=\max\left(\mathcal{R}^r\left(\mathbf{a}_1\right)-\log_2\left(|\lambda|\right),0\right),
\end{split}
\end{equation}
which concludes the proof, since $\log_2\left(|\lambda|\right)>0$.
\end{IEEEproof}
\begin{corollary}\label{Corollary1}
Between all collinear candidate equation coefficient vectors, the
one achieving the best computation rate is the one whose elements
have their greatest common divisor (GCD) equal to $1$.
\end{corollary}
\begin{IEEEproof}
If $\mathbf{a}_1$ is a vector with elements with GCD equal to $1$, every other collinear vector
$\mathbf{a}_2$ can be written as
$\mathbf{a}_2=\lambda\mathbf{a}_1$, where $\lambda$ is the
GCD of the elements of $\mathbf{a}_2$, while its
sign depends on whether $\mathbf{a}_1$ and $\mathbf{a}_2$ have the
same or opposite direction and $|\lambda|>1$. Thus, according to
Lemma 1, $\mathbf{a}_1$ achieves greater computation rate. This
concludes the proof.
\end{IEEEproof}

The complexity of the proposed algorithm depends on the length of sets $\Omega_m$. Thus, the termination of adding vectors to the candidate equation coefficient vector set $\Omega_m$ is crucial and may be chosen according to different criteria. We concentrate on three cases: i)guaranteeing the existence of at least one full rank matrix, ii) satisfying specific minimum rate requirements $V_l$ for each source $l$, iii) satisfying a complexity / performance optimization trade-off of a practical system. In the first case, the selection of candidates terminates when the sets $\Omega_m$ span the $L$-dimensional space for the first time, since at that case, at least one full rank matrix $\mathbf{A}$ exists. In the second case, the addition of vectors terminates when it no longer holds $\mathcal{R}^r_m\geq V_l$ when $a_{m,l}\neq 0$, that is, the achievable computation rates satisfy the constraints set by each user. In the third case, the size of the sets is defined by the complexity constraints of a practical system. Therefore, a smaller size of sets can be selected in order to reduce the complexity, at the expense of the network performance.

After the termination of adding vectors to the sets $\Omega_m$, we delete all the vectors which have a GCD different than $1$, (implementation of Corollary \ref{Corollary1}), and keep one of the two vectors, if a pair of opposite vectors appears in $\Omega_m$. According to the candidate set $\Omega_m$, each relay also constructs the set of corresponding achievable computation rates, denoted by $\Gamma_m$.

\vspace{-0.15in}
\subsection{Coalition formation strategy}\label{SS:Strategy}

The proposed strategy aims to find the best partition of players, containing the top-coalition $\mathcal{S}^*$ of $L$ players and the CD, as described in Theorem \ref{core_partition}. Note that, such a coalition achieves the best non-zero utility and the corresponding partition $\pi^*$ is in the core, as proven in Theorem \ref{core_partition}, so in the following, the term coalition is used to denote the top-coalition $\mathcal{S}^*$ of $L$ relays cooperating with the CD.

In the proposed algorithm, the players sort their achievable computation rates in descending order in a list $Q$, so that the vectors achieving the best rates are used first, as done in \cite{Wei2} and in the greedy algorithm of \cite{Caire2}. However, it continues searching for solutions on the Pareto frontier, instead of stopping when the minimum rate is maximized. The strategy of the players is described as follows:
\begin{enumerate}
\item The initial partition consists of the coalition which contains the CD and $L$ relays and also maximizes the minimum rate. A modified version of the algorithm proposed in \cite{Wei2} is used. As explained, the solution offered by this coalition may not be a point on the Pareto frontier.
\item The players try to form a coalition with the CD which a) contains the relay which achieved the minimum computation rate in the coalition of step (1), so that the minimum rate is unchanged, b) contains $L-1$ relays achieving higher computation rates than the aforementioned relay, c) achieves higher sum rate than the coalition in step (1), and the best sum rate among all coalitions with the same minimum rate. Dominated solutions are excluded from this search, with the use of a weight function for each candidate vector, described in the next subsection. Thus, only vectors with a suitable weight are considered.
\item If such a coalition as described in step (2) can be found, then the coalition of step (1) deforms, and the coalition in step (2) forms. Note that in this case, the new coalition corresponds to a point on the Pareto frontier. If no such coalition can be found, then the coalition of step (1) corresponds to a point on the Pareto frontier.
\item The players now check if they can form a coalition which decreases the minimum rate but increases the sum rate. Again, only vectors with suitable weight are considered. Note that, since the QoS requirements of the users are a priori satisfied when the vector sets are constructed, the minimum rate can be gradually reduced with a search among the sorted list $Q$.
\item If such a coalition can be found, the previous coalition deforms and the coalition of step (4) forms. The new coalition corresponds to a new point on the Pareto frontier which achieves a higher sum rate.
\item Steps (4) and (5) are repeated until no other coalition with higher sum rate can be found. The stable partition contains the coalition which corresponds to the Point on the Pareto frontier with the highest sum rate.
\end{enumerate}

The above procedure can be summarized in Algorithm 1.

\begin{algorithm}\label{Alg:split_merge}
\caption{: Coalition Formation}
\begin{algorithmic}[1]
\State \underline{\textbf{Step 1}} The players form a partition which contains a coalition of $L$ relays and the CD, with corresponding coding matrix $\mathbf{A}_1$, for which:
\State \quad $\mathbf{A}_1=\mathrm{arg}\max\limits_{|\mathbf{A}|\neq 0}\left(\min\limits_{m:\mathbf{a}_m\in\mathbf{A}}R_m^r\right)$.
\State \underline{\textbf{Step 2}} The players search for a new partition containing a coalition with coding matrix $\mathbf{A}_2$, for which:
\State \quad $\min\limits_{m:\mathbf{a}_m\in\mathbf{A}_1}R_m^r=\min\limits_{m:\mathbf{a}_m\in\mathbf{A}_2}R_m^r$ and
\State \quad $\not\exists i: \left(\min\limits_{m:\mathbf{a}_m\in\mathbf{A}_2}R_m^r=\min\limits_{m:\mathbf{a}_m\in\mathbf{A}_i}R_m^r\right)\cap\left(\sum\limits_{\begin{subarray}{c}l=1\\\mathbf{a}_m\in\mathbf{A}_2\end{subarray}}^L\mathcal{R}_l^s<\sum\limits_{\begin{subarray}{c}l=1\\\mathbf{a}_m\in\mathbf{A}_i\end{subarray}}^L\mathcal{R}_l^s\right)$.
\State \underline{\textbf{Step 3}}\If {there is such a coalition with coding matrix $\mathbf{A}_2$,} the coalition of Step 1 deforms and the coalition of Step 2 forms, while the output matrix of this step is $\mathbf{A}_3=\mathbf{A}_2$. \Else, $\mathbf{A}_3=\mathbf{A}_1$.\EndIf \State After this step, the current coalition corresponds to a point on the Pareto frontier, with coding matrix $\mathbf{A}_3$.
\State \underline{\textbf{Step 4}} The players search for a new partition containing a coalition with coding matrix $\mathbf{A}_4$, for which:
\State \quad $\min\limits_{m:\mathbf{a}_m\in\mathbf{A}_4}R_m^r<\min\limits_{m:\mathbf{a}_m\in\mathbf{A}_3}R_m^r$,
\quad $\sum\limits_{\begin{subarray}{c}l=1\\\mathbf{a}_m\in\mathbf{A}_4\end{subarray}}^L\mathcal{R}_l^s\geq \sum\limits_{\begin{subarray}{c}l=1\\\mathbf{a}_m\in\mathbf{A}_3\end{subarray}}^L\mathcal{R}_l^s$,
\quad and
\State \quad $\not\exists i:\left(\min\limits_{m:\mathbf{a}_m\in\mathbf{A}_4}R_m^r=\min\limits_{m:\mathbf{a}_m\in\mathbf{A}_i}R_m^r\right)\cap\left(\sum\limits_{\begin{subarray}{c}l=1\\\mathbf{a}_m\in\mathbf{A}_4\end{subarray}}^L\mathcal{R}_l^s<\sum\limits_{\begin{subarray}{c}l=1\\\mathbf{a}_m\in\mathbf{A}_i\end{subarray}}^L\mathcal{R}_l^s\right)$.
\State \underline{\textbf{Step 5}}\If {there is such a coalition with coding matrix $\mathbf{A}_4$,} the coalition of Step 3 deforms and the coalition of Step 4 forms, while the output matrix of this step is $\mathbf{A}_5=\mathbf{A}_4$. Steps 4 and 5 are repeated with new $\mathbf{A}_3=\mathbf{A}_5$, until the algorithm stops. \Else, $\mathbf{A}_5=\mathbf{A}_3$ and the algorithm stops.\EndIf \State After this step, the current coalition corresponds to the point on the Pareto frontier, which achieves the highest sum rate.
\end{algorithmic}
\end{algorithm}

\subsection{Elimination of dominated solutions: the weight function}\label{SS:weight}
During the process of forming coalitions, each relay has many candidate coefficients that it can use. However, some of them may a priori be excluded from the search, since they cannot lead to a higher sum rate. In order to exclude these dominated solutions, we introduce a weight function which computes an ideal sum rate for each vector, based on the vectors selected by other relays which are already in a coalition, plus the best possible choice of vectors for the relays that will enter the coalition afterwards.

More specifically, this
weight is constructed by
i) the already defined transmission rates due to other vectors in
the candidate coalition, ii) the
transmission rates which are defined by the specific candidate vector for which the weight is computed, and
iii) the best possible transmission rates that can be achieved in the
current list $Q$ of all computation rates of the relays, for all those sources with still undetermined
transmission rates. The function which computes the weight for a
candidate vector $\mathbf{a}$ is given in Algorithm 2. Matrix $\mathbf{A}_{L-l}$ contains all vectors of players which are already in the candidate coalition, while $l$ indicates the number of relays which are needed, so that a coalition will be of $L$ players and the CD.

Note that the use of weights is a strategy of elimination of dominated solutions, so that an exhaustive search is not needed. The possible combinations of vectors for a full rank matrix $\mathbf{A}$ are drastically reduced since there is no
need to search all the elements of the sorted list $Q$.

\begin{algorithm}\label{Alg:weight}
\caption{: $\mathrm{weight}(l,Q,\mathbf{A}_{L-l},\mathbf{a})$}
\begin{algorithmic}[1]
\State The player with candidate vector $\mathbf{a}$ computes the
following weight \State \textbf{i}. The player checks the vectors
in the matrix $\mathbf{A}_{L-l}$ and computes the transmission
rates that are already defined by this matrix, using
(\ref{transmit_rate_2}). \State \textbf{ii}. The player computes
the transmission rates which are determined by the vector
$\mathbf{a}$. \State \textbf{iii}. If there are still undetermined
transmission rates, they are computed according to the vector in
$Q$ with the maximum possible rate, which has non-zero
corresponding element and belongs to the candidate set of a relay
which is not the current player or does not contribute to the matrix
$\mathbf{A}_{L-l}$, that is, it is not in the candidate coalition. \State \textbf{iv}. The weight of $\mathbf{a}$
is the sum of the above determined transmission rates.
\end{algorithmic}
\end{algorithm}

\section{Illustrative Examples}\label{S:Examples}
\subsection{Pareto Frontier and Pareto Optimal Point
Examples}\label{SS:pareto_examples}

In this section we present two examples concerning a network with
$L=4$ sources and $M=5$ relays, for two random channel instances and $P=10$ dB.
These examples present the solution sets containing all possible
full rank matrices $\mathbf{A}$, and the minimum and sum rate that
each solution achieves.

\begin{figure}[t!]
\centering\includegraphics[keepaspectratio,width=0.45\columnwidth]{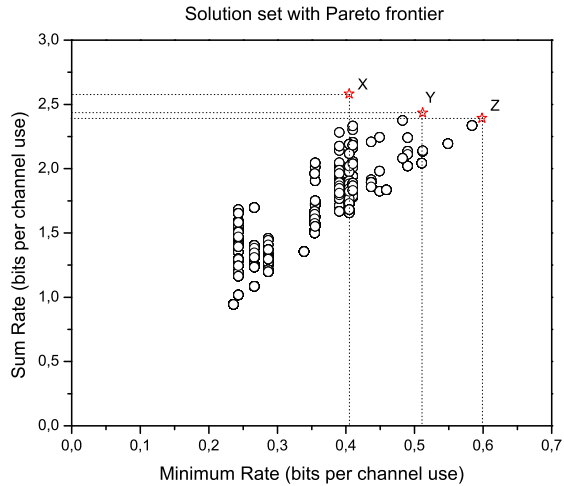}
\caption{Study case for a network with $L=4$ sources and $M=5$
relays for a channel realization which leads to a solution set
with Pareto frontier.} \label{Fig:Pareto_frontier}
\end{figure}

In Fig. \ref{Fig:Pareto_frontier}, the Pareto frontier consists of
three distinct points, marked with asterisks. These solutions
illustrate the tradeoff between the maximization of the minimum
rate and the maximization of the sum rate. We consider that the minimum rate requirement of all the users is $0.2359$ bits/Hz/Channel use\footnote{This choice corresponds to the value which ensures that all sets $\Omega_m$ span the $L$-dimensional space and thus at least one full rank matrix $\mathbf{A}$ exists.}. Note that, after Step 2
of the coalition formation algorithm is applied, the coalition which forms is
the one which corresponds to point Z. When Step 4 is applied for the first time,
the coalition which forms
corresponds to point Y, while after the second time Step 4 is applied, the coalition which forms corresponds to point
X and  the algorithm stops.

\begin{figure}[t!]
\centering\includegraphics[keepaspectratio,width=0.45\columnwidth]{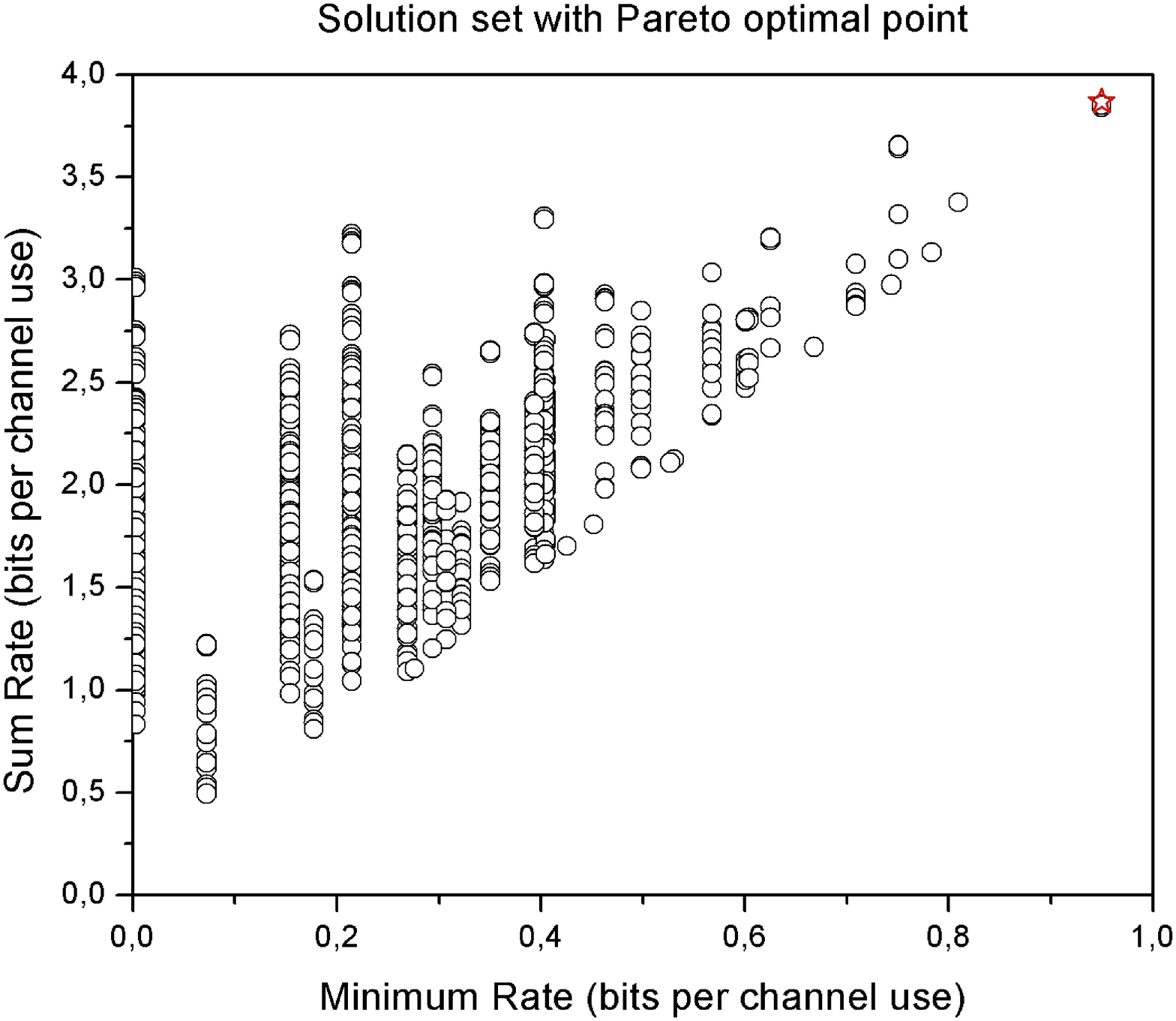}
\caption{Study case for a network with $L=4$ sources and $M=5$
relays for a channel realization which leads to a solution set
with Pareto optimal point.} \label{Fig:Pareto_optimal_point}
\end{figure}

In the second example, depicted in Fig.
\ref{Fig:Pareto_optimal_point}, the solution set contains a Pareto
optimal point, marked with asterisk. We consider that the users do not have specific rate requirements, so they are set to zero. Note that, as shown in Fig.
\ref{Fig:Pareto_optimal_point}, apart form the Pareto optimal
point, there is also another point which maximizes the
minimum rate, but does not achieve the maximum sum rate. In this
case, when Step 1 of the algorithm is applied, any of these two points
may be chosen, since the optimization is
performed according to the maximization of the minimum rate. However, when Step 2 is
also applied, it leads to the best possible solution, both in
terms of minimum rate and sum rate, which is the Pareto optimal
point.

\subsection{Transparent Realization}\label{SS:transparent_realization}

In this section we present a numerical example of the
implementation of the proposed algorithms. We consider a network
with $L=4$ transmitters and $M=5$ relays, operating with power constraint $P=10$ dB. This example corresponds
to the solution set depicted in Fig. \ref{Fig:Pareto_frontier}, where $V_l=V=0.2359$ bits/Hz/channel use.
The optimization is performed for the following channel
coefficient vectors for each relay respectively:
\begin{equation}\label{channel_example}
\begin{split}
&\mathbf{h}_1=\left[1.1408\,\,0.9331\,\,-0.5206\,\,-0.5897\right]^T,\mathbf{h}_2=\left[-0.8927\,\,0.9095\,\,0.3323\,\,0.2708\right]^T,\\
&\mathbf{h}_3=\left[-0.6112\,\,1.1819\,\,0.6595\,\,-0.7272\right]^T,\mathbf{h}_4=\left[-1.0152\,\,-0.4052\,\,-0.5168\,\,-0.2829\right]^T,\\
&\mathbf{h}_5=\left[-1.1169\,\,-0.8165\,\,-0.4853\,\,0.6650\right]^T.
\end{split}
\end{equation}\vspace{-0.1in}
The sets $\Omega_m$ which are constructed as described in Section
\ref{SS:candidate_sets} are given below:
\begin{equation}\label{Omega_example}
\begin{split}
&\Omega_1=\left[\begin{smallmatrix}2&1&1&2&1\\2&1&1&1&1\\-1&-1&0&-1&0\\-1&-1&0&-1&-1\end{smallmatrix}\right],\quad
\Omega_2=\left[\begin{smallmatrix}-1&0&-1&-1&-2\\1&1&0&1&2\\0&0&0&1&1\\0&0&0&0&1\end{smallmatrix}\right],\quad\Omega_3=\left[\begin{smallmatrix}-1&-1&0&0&0\\2&1&1&1&1\\1&1&0&1&0\\-1&-1&0&-1&-1\end{smallmatrix}\right],\\
&\Omega_4=\left[\begin{smallmatrix}-1&-1&-2&-2\\0&0&-1&-1\\0&-1&-1&-1\\0&0&-1&0\end{smallmatrix}\right],\quad\Omega_5=\left[\begin{smallmatrix}-2&-1&-1&-1&-1\\-1&-1&-1&0&-1\\-1&-1&0&0&0\\1&1&1&0&0\end{smallmatrix}\right].
\end{split}
\end{equation}

The corresponding $\Gamma_m$ sets containing the achievable rates
are
\begin{equation}\label{Gamma_example}
\begin{split}
&\Gamma_1=[0.5984\,\,0.5107\,\,0.4825\,\,0.4588\,\,0.4367],\Gamma_2=[0.8742\,\,0.4100\,\,0.3901\,\,0.3544\,\,0.2359],\\
&\Gamma_3=[0.8989\,\,0.6047\,\,0.4897\,\,0.2866\,\,0.2430],\Gamma_4=[0.7127\,\,0.4047\,\,0.3389\,\,0.2663],\\
&\Gamma_5=[0.5839\,\,0.5486\,\,0.5119\,\,0.4490\,\,0.3549].
\end{split}
\end{equation}

If $\boldsymbol{\omega}_{m,i}$ is the $i$-th column of $\Omega_m$ and and
$\gamma_{m,i}$ is the $i$-th element of $\Gamma_m$, then the
sorted list $Q$ is the following:
\begin{equation}\label{list_Q}
\begin{split}
Q=\{&\gamma_{3,1},\gamma_{2,1},\gamma_{4,1},\gamma_{3,2},\gamma_{1,1},\gamma_{5,1},\gamma_{5,2},\gamma_{5,3},\gamma_{1,2},\gamma_{3,3},\gamma_{1,3},\gamma_{1,4},\gamma_{5,4},\gamma_{1,5},\gamma_{2,2},\gamma_{4,2},\gamma_{2,3},\\
&\left.\gamma_{5,5},\gamma_{2,4},\gamma_{4,3},\gamma_{3,4},\gamma_{4,4},\gamma_{3,5},\gamma_{2,5}\right\}.
\end{split}
\end{equation}

When Step 1 of the algorithm is executed, the coalition which forms corresponds the full rank matrix
\begin{equation}\label{output_Algorithm_1}
\mathbf{A}_1=\left[\begin{smallmatrix}-1&-1&-1&2\\2&1&0&2\\1&0&0&-1\\-1&0&0&-1\end{smallmatrix}\right].
\end{equation}
In this case, the relays which form the coalition are $R_3,R_2,R_4$ and $R_1$, using the coding
vectors $\boldsymbol{\omega}_{3,1},\boldsymbol{\omega}_{2,1},\boldsymbol{\omega}_{4,1}$ and
$\boldsymbol{\omega}_{1,1}$. Using (\ref{transmit_rate_2}), every transmission
rate is defined by $\gamma_{1,1}$, thus the minimum rate over the
network is $
\min\left(\mathcal{R}^s_l\right)=\gamma_{1,1}=0.5984$ while the sum rate is
$\sum_{l=1}^4\mathcal{R}^s_l=4\gamma_{1,1}=2.3936$.
When Step 2 is executed, no other matrix with the same minimum rate but better sum rate can be found, so in Step 3 it is $\mathbf{A}_3=\mathbf{A}_1$.

Now let the minimum rate requirement be relaxed. When Step 4 is executed, it first tries to replace $\boldsymbol{\omega}_{1,1}$ in matrix $\mathbf{A}$ with another vector with $\gamma_{m,i}<\gamma_{1,1}$. If $W\left(\boldsymbol{\omega}_{m,i}\right)$ denotes the weight of $\boldsymbol{\omega}_{m,i}$ according to the list $Q$, the following weights are calculated:
\begin{itemize}
\item For $\gamma_{5,1}$, since there is no zero element, $W\left(\boldsymbol{\omega}_{5,1}\right)=4\gamma_{5,1}=2.3356$, which is smaller than the previously calculated sum rate, thus $\boldsymbol{\omega}_{5,1}$ is discarded.
\item The next rate in $Q$ is $\gamma_{5,2}$ and since there is no zero element, $W\left(\boldsymbol{\omega}_{5,1}\right)=4\gamma_{5,2}=2.1944$, which is again smaller than the previously calculated sum rate, thus $\boldsymbol{\omega}_{5,2}$ is discarded.
\item For $\gamma_{5,3}$, there is one zero element in $\boldsymbol{\omega}_{5,3}$, thus the corresponding transmission rate, $\mathcal{R}^s_3$ will be assigned an ideal value, the best possible according to the list $Q$, which is $\gamma_{3,1}$. Thus, $W\left(\boldsymbol{\omega}_{5,3}\right)=3\gamma_{5,3}+\gamma_{3,1}=2.4346$, which is better than the previously calculated sum rate. Thus $\boldsymbol{\omega}_{3,1}$ is chosen.
\end{itemize}
After Step 4, a coalition with the following full rank matrix is found:
\begin{equation}\label{output_Algorithm_2}
\mathbf{A}_4=\left[\begin{smallmatrix}-1&-1&-1&-1\\2&1&0&-1\\1&0&0&0\\-1&0&0&1\end{smallmatrix}\right],
\end{equation}
which contains the vectors $\boldsymbol{\omega}_{3,1}$, $\boldsymbol{\omega}_{2,1}$, $\boldsymbol{\omega}_{4,1}$ and $\boldsymbol{\omega}_{5,3}$.
The minimum rate is $\min\left(\mathcal{R}^s_l\right)=\gamma_{5,3}=0.5119$
while the sum rate is $\sum_{l=1}^4\mathcal{R}^s_l=3\gamma_{5,3}+\gamma_{3,1}=2.4346$.
Note that the actual sum rate in this case coincides with $W\left(\boldsymbol{\omega}_{5,3}\right)$, while the minimum rate meets the QoS requirement. The matrix in (\ref{output_Algorithm_2}) corresponds to the solution point Y in Fig. \ref{Fig:Pareto_frontier}, and so after Step 5, $\mathbf{A}_5=\mathbf{A}_4$. Steps 4 and 5 are repeated in similar manner, and the algorithm ends with the following coding matrix as a final solution:
\begin{equation}\label{output_final}
\mathbf{A}=\left[\begin{smallmatrix}-1&-1&-1&-1\\2&1&0&0\\1&0&0&-1\\-1&0&0&0\end{smallmatrix}\right],
\end{equation}
which contains the vectors $\boldsymbol{\omega}_{3,1}$, $\boldsymbol{\omega}_{2,1}$, $\boldsymbol{\omega}_{5,4}$ and $\boldsymbol{\omega}_{4,2}$.
The minimum rate is $\min\left(\mathcal{R}^s_l\right)=\gamma_{4,2}=0.4047$
while the sum rate is $\sum_{l=1}^4\mathcal{R}^s_l=2\gamma_{4,2}+\gamma_{2,1}+\gamma_{3,1}=2.5825$.
Note that the selection of the matrix $\mathbf{A}$ is the dominant strategy, since it corresponds to the point on the Pareto frontier which maximizes the sum rate.

The complexity of the algorithm generally increases when the length of $\Omega_m$, the value of $L$ and the value of $M$ increase. However, the use of the weight function, combined with the specific order of the search among vectors, using the list $Q$, drastically reduces the number of coding matrices, whose sum rate and minimum rate are actually computed by the algorithm. This reduction is evident for this transparent realization in Table \ref{Table:Algorithms}. More specifically, an exhaustive search among all possible combinations of vectors between different relays would require the formation of 2625 matrices for this specific example, while only 1809 of them are full rank. However, with the use of the proposed algorithm and the use of the weight function, only 12 matrices are actually formed and their sum rate is checked, while all the rest are excluded since they are dominated solutions. Finally, from these 12 matrices, only 3 correspond to points on the Pareto frontier, and thus only 3 coalitions actually form.

\begin{table}[t!]
\centering \caption{Formation of matrices by the coalition formation algorithm}\label{Table:Algorithms}
\begin{tabular}{|l||c|}
\hline \textbf{Formation of Matrices and Sum rates} &\textbf{Number of Coding Matrices}\\
\hline
All combinations of vectors & $2625$\\
\hline
All full rank matrices & $1809$\\
\hline
Matrix searches performed by Steps 2 and 4 & $12$\\
\hline
Coalition formations performed by Steps 3 and 5 & $3$\\
\hline
\end{tabular}
\end{table}


\vspace{-0.3in}
\section{Simulation Results}\label{S:Simulation_results}
In the following, we present simulation results for the proposed
coalition formation strategy, for Gaussian distributed channel
coefficients with distribution $\mathcal{N}\left(0,1\right)$. The results examine the effect of available relays on the sum rate and the utility functions of the players.

\begin{figure}[h!]
\centering\includegraphics[keepaspectratio,width=0.48\columnwidth]{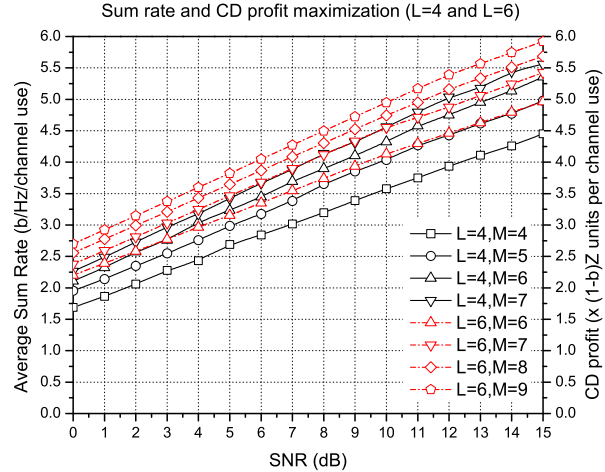}
\caption{Average Sum Rate and Average CD profit, for L=4, M=4,5,6,7, and L=6, M=6,7,8,9.} \label{Fig:CD_profit}
\end{figure}

In Fig. \ref{Fig:CD_profit}, we study the effect of relay selection on the sum rate and profit maximization. The sum rate of a network with $L=4$ users and $M=4,5,6,7$ relays (black lines) and a network with $L=6$ users and $M=6,7,8,9$ relays (red lines) is depicted. The profit of the CD is also indicated for arbitrary values of $Z$ and $(1-b)$, according to eqs. (\ref{revenue2}) and (\ref{revenue_players}). The sets used for the candidate vectors of the relays, are those constructed as described in Section \ref{SS:candidate_sets} when the full rank property is guaranteed. Simulations show that relay selection significantly improves the network performance in terms of sum rate and profit. For $L=4$, when the number of available relays is increased from $M=4$ to $M=5$, relay selection improves the performance of the network up to $3$ dB, for $SNR=15$ dB, while when the number of relays is increased to $M=7$, the gain is up to $6$
dB, compared to the case of no relay selection ($M=4$). Moreover, from Fig. \ref{Fig:CD_profit} one can observe that, the gain is smaller in each addition of an extra relay, compared to the previous addition. Similar conclusions are shown for the case of $L=6$. This implies that as the number $M$ of relays increases, although there is no upper bound to the performance, the improvement due to relay selection becomes negligible, and there is no reason to add extra nodes to the network.

\begin{figure}[h!]
\centering\includegraphics[keepaspectratio,width=0.48\columnwidth]{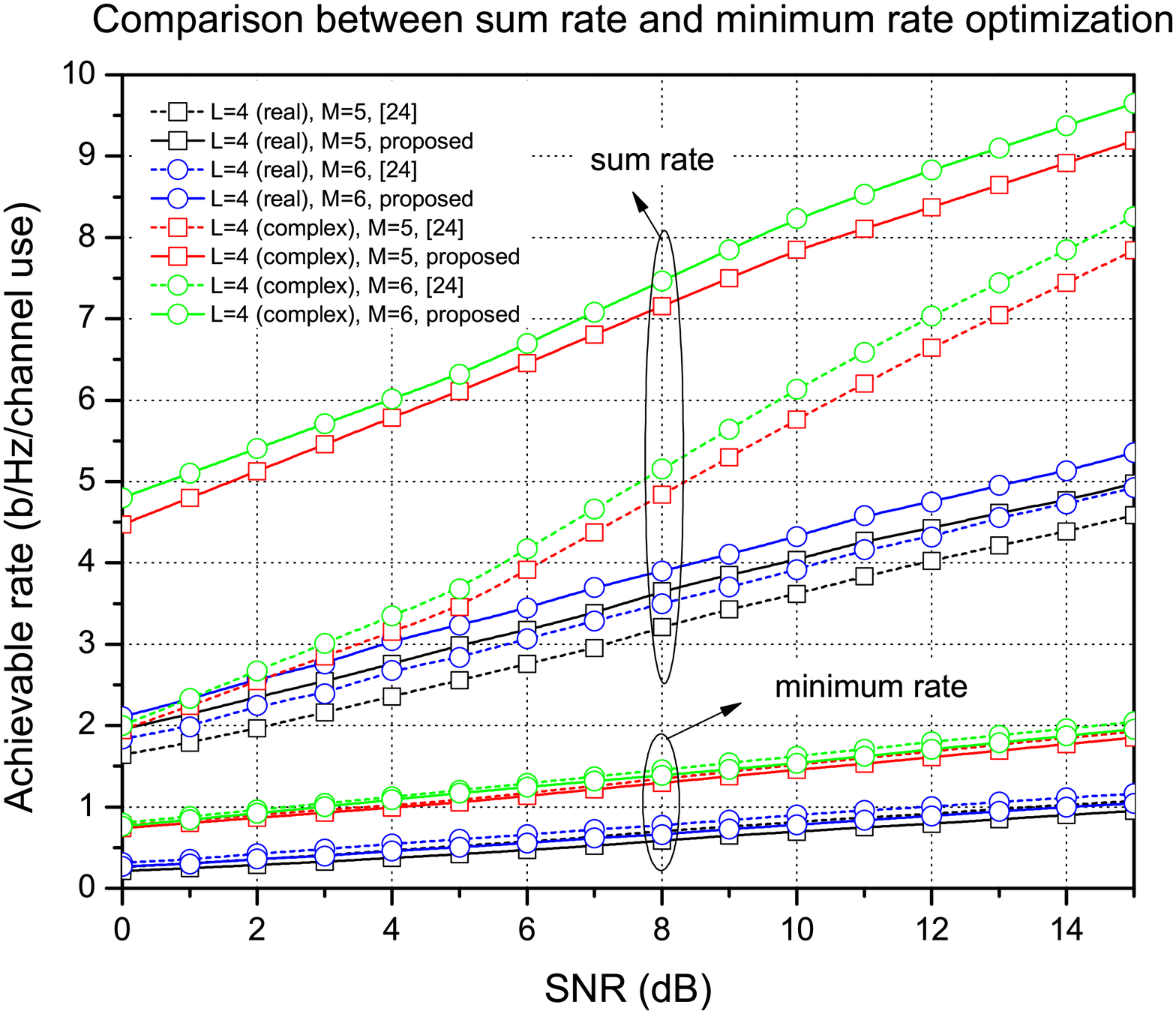}
\caption{Average Sum Rate and Average minimum rate for the optimization in \cite{Wei2} and the proposed strategy, for L=4 and M=5,6 (real and complex signaling).} \label{Fig:comparison}
\end{figure}

\begin{figure}[h!]
\centering\includegraphics[keepaspectratio,width=0.48\columnwidth]{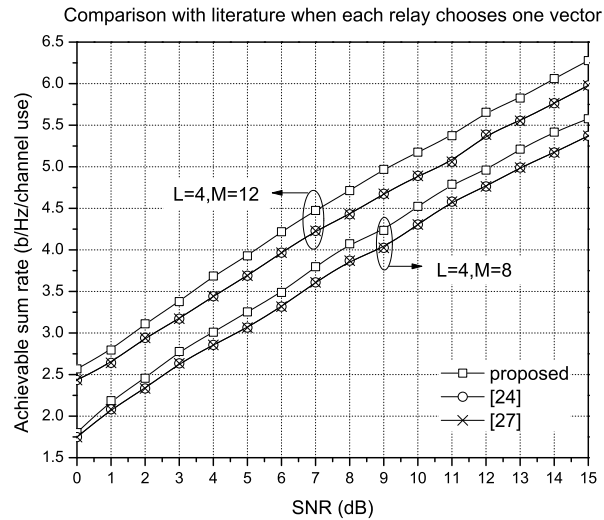}
\caption{Average Sum Rate for the optimization in \cite{Wei2}, \cite{Caire2} and the proposed strategy, for L=4 and M=8,12, when each relay chooses only one vector.} \label{Fig:comparison27}
\end{figure}

\begin{figure}[h!]
\centering\includegraphics[keepaspectratio,width=0.48\columnwidth]{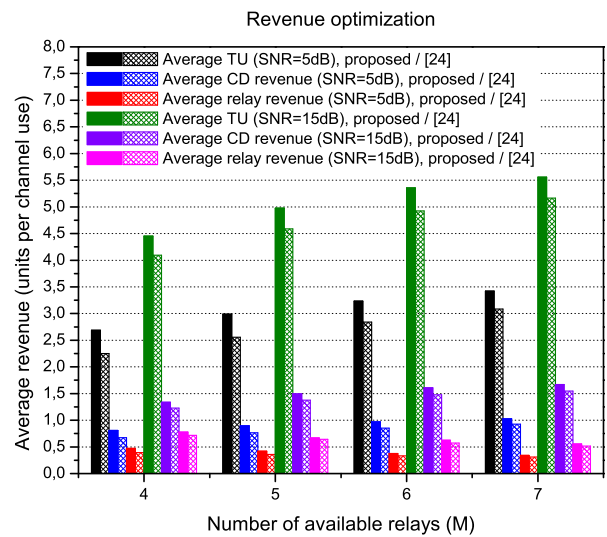}
\caption{Revenue comparison between the optimization in \cite{Wei2} (pattern) and the proposed strategy (solid color), for L=4 and M=4,5,6,7.} \label{Fig:revenue}
\end{figure}

In Fig. \ref{Fig:comparison}, the sum rate improvement offered by the proposed optimization is illustrated, compared to the optimization performed in \cite{Wei2}, which maximizes the minimum transmission rate\footnote{The results of \cite{Wei2} are calculated according to the logarithm with base 2 in this work, instead of the natural logarithm which was used in \cite{Wei2}.}. The minimum transmission rate is also illustrated. The presented results are for a network with $L=4$ sources and
$M=5,6$ relays, with real or complex signaling. For a cost in the average minimum transmission
rate, a gain of up to $3$ dB in the sum rate is achieved for real signaling, for $M=5$, while a gain of up to $2.5$  dB is achieved for $M=6$. It is remarkable that the proposed strategy for $M=5$ outperforms the sum rate achieved by the optimization in \cite{Wei2} even for $M=6$, which means that an algorithmic optimization, which is performed by software, may be favorable, compared to adding an extra node to the network, leading to reduced overall cost. The gains for complex signaling are even greater, especially for low SNR values, reaching a value of up to $6$ dB. This is because the proposed optimization involves both the real and imaginary part of the candidate vectors. Note that the optimization of a complex system with $L=4$ corresponds to a real system of $L'=8$.

In Fig. \ref{Fig:comparison27}, we present a special case when each $\Omega_m$ comprises only one vector, in order to directly compare with the scheme of \cite{Caire2}. The sum rate of the proposed algorithm for $L=4$ and $M=8,12$ with real signaling is compared to the optimization schemes in \cite{Wei2,Caire2}. In our work, since we compare achievable rates when $N\rightarrow\infty$, we consider the case of a field size $p\rightarrow\infty$ \cite{Nazer}, which leads to the existence of few zero elements in the coding matrix. Thus, a network decomposition is seldom achieved by the strategy in \cite{Caire2}, which can also be seen in the figure, since its performance is almost identical to that in \cite{Wei2}. Since the proposed algorithm allows non-symmetric transmission rates, it achieves a gain up to $1.5$ dB for the illustrated cases. Note that, if the zero elements allow a network decomposition, the execution of our algorithm will also perform the decomposition, in addition to exploiting the gain offered by non-symmetric rates.

In Fig. \ref{Fig:revenue}, the average total utility of the top-coalition $\mathcal{S}^*$, the average revenue for the CD and the average revenue for one relay is depicted, for the case of $L=4$ and for $M=4,5,6,7$, when $SNR=5,15$ dB. The solid color corresponds to the proposed optimization while the pattern corresponds to the optimization in \cite{Wei2}. The sets $\Omega_m$ are constructed in a way to guarantee the full rank property. The parameters for the revenues which were used are $Z=1$ and $b=0.7$. The gain over previous results is up to $21\%$ for the TU and for the case of $M=4$, while the results are similar for other values of $M$. Furthermore, both the profit of the CD and the average profit of each relay are significantly increased with the proposed strategy. Another important observation is that the TU and the CD profit increase, when the number of available relays increases. However, for both optimization techniques, the average profit of one relay reduces. This is expected, since the relay is not operating at all times (i.e., it is not always in the top-coalition $\mathcal{S}^*$), and thus it receives zero profit when it does not operate. Note that, in a practical scenario, the relays which are not selected for a specific channel realization and do not receive profit at that instant, may serve another group of users which do not belong to this network, increasing their total profit.

\vspace{-0.3in}
\section{Conclusions}\label{S:Conclusions}
In this paper, a cloud network with surplus relays with respect to the users, which employs Compute-and-Forward, was studied. Relay selection and PNC at the relays were performed, which lead to the maximization of the total throughput under minimum rate constraints. This maximization was explored with the use of the Pareto frontier, while the relay selection was matched to a coalition formulation game. An efficient algorithm for the coalition formation was proposed, which leads to the optimal solution in terms of sum rate and players' profit. Extensive examples and simulation results showed that the proposed strategy leads to an improvement of at least $1.5$ dB for the sum rate of the network, compared to previous results in the literature. Furthermore, the profit of the players is increased up to $21$ \%, while the solution provided by the proposed algorithm has no profitable deviation, guaranteeing the stability of the partitioning of the players.



\vspace{-0.15in}
\appendices
\section{Proof of Theorem \ref{stable_game}}

According to \cite{Banerjee}, given a non-empty set of players $\mathcal{V}\subseteq\mathcal{N}$, a non-empty subset of players $\mathcal{S}\subseteq \mathcal{V}$ is a top-coalition of V, iff for any $i\in\mathcal{S}$ and any $\mathcal{T}\subseteq \mathcal{V}$ with $i\in \mathcal{T}$, we have $\mathcal{S}\succeq_i \mathcal{T}$. A coalition formation game satisfies the top-coalition property iff for any non-empty subset of players $\mathcal{V}\subseteq\mathcal{N}$, there exists a top coalition of $\mathcal{V}$. Next, we prove that the proposed game has a non-empty core.

Consider player $i \in \mathcal{N}$. For any $\mathcal{S}, \mathcal{T} \in \mathcal{S}_i(\mathcal{N})$ with $\mathcal{S} \neq \mathcal{T}$, we define the following preference, using the profit of each player as defined in eq. (\ref{revenue_players}):
\begin{equation}
\mathcal{S}\succeq_i \mathcal{T}\Longleftrightarrow \phi_i(\mathcal{S})\geq\phi_i(\mathcal{T})
\end{equation}
We next show that there is a top coalition for each $\mathcal{V}\subseteq\mathcal{N}$ with $\mathcal{V}\neq0$. Let
$\mathcal{Q(V)}=\{\mathcal{S}\in 2^\mathcal{V}\backslash\{\emptyset\}: \phi_i(\mathcal{S})\geq\phi_i(\mathcal{T}) \,\forall\,\mathcal{T}\in 2^\mathcal{V}\backslash\{\emptyset\} \textit{, }\forall i\in \mathcal{S}\}$.
\begin{itemize}
\item If $\mathcal{V}\ni\mathcal{C}$ and $|\mathcal{V}|\geq L+1$, according to eq. (\ref{revenue2}), there is always at least one coalition with non-zero TU. The revenue of the CD is proportional to the TU, while the revenue of the relays is proportional to the TU and inverse proportional to the size of the coalition. Thus, the set $\mathcal{Q}(\mathcal{V})$ is non empty, containing the coalitions with minimum size, i.e. $|\mathcal{S}|=L+1$, which maximize the TU.
\item In any other case, the TU and the revenue of the players is zero, thus the condition $\phi_i(\mathcal{S})\geq\phi_i(\mathcal{T})$ holds with equality for any arbitrary non-empty set $\mathcal{Q}(\mathcal{V})$.
\end{itemize}
Consequently, each element $\mathcal{S}$ of $\mathcal{Q(V)}$ is a top-coalition of $\mathcal{V}$. This is because, for any $i\in \mathcal{S}$ and any $\mathcal{T} \in 2^\mathcal{V}\backslash\{\emptyset\}$ we have $\phi_i(\mathcal{S})\geq\phi_i(\mathcal{T})$ which denotes a preference $\mathcal{S}\succeq_i \mathcal{T}$. Therefore the top coalition property is satisfied and thus there is a core partition, as shown in \cite{Banerjee}.
\vspace{-0.15in}
\section{Proof of Theorem \ref{core_partition}}

Let $\mathcal{V}_0=\mathcal{N}$, then $\mathcal{S}^*$ is a top-coalition of $\mathcal{V}_0$. That is, for any $i\in\mathcal{S}^*$ and any $T\subseteq\mathcal{V}_0$ with $i\in\mathcal{T}$, $\mathcal{S}^*\succeq_i\mathcal{T}$. Indeed, for all $i\in\mathcal{S}^*$, $\phi_i(\mathcal{S}^*)\geq\phi_i(\mathcal{T})$, for any $\mathcal{T}\subseteq\mathcal{V}_0$. This is because, when $|\mathcal{T}|\geq L+1$ and the coalition $\mathcal{T}$ contains the CD, then the sum rate of $\mathcal{T}$ cannot be greater than that of $\mathcal{S}^*$ and thus, according to eq. (\ref{revenue2}) and eq. (\ref{revenue_players}), $\phi_i(\mathcal{S}^*)\geq\phi_i(\mathcal{T})$. In any other case, according to eq. (\ref{revenue2}) and eq. (\ref{revenue_players}), $\phi_i(\mathcal{T})=0\leq\phi_i(\mathcal{S}^*)$.

Now let $\mathcal{V}_1=\mathcal{V}_0\backslash\{\mathcal{S}^*\}$, then $\mathcal{S}_1$ is a top-coalition of $\mathcal{V}_1$. Since $|\mathcal{S}_1|=1$, then if $i\in\mathcal{S}_1$, $\phi_i(\mathcal{S}_1)=0$. However, since $\mathcal{C}\not\in\mathcal{V}_1$, $\phi_i(\mathcal{T})=0$ for any $\mathcal{T}\subseteq\mathcal{V}_1$ and thus, $\phi_i(\mathcal{S}_1)\geq\phi_i(\mathcal{T})$ for any $\mathcal{T}\subseteq\mathcal{V}_1$ (which actually holds with equality). Thus $\mathcal{S}_1$ is a top-coalition of $\mathcal{V}_1$. Working in the same way for $\mathcal{S}_2,\ldots,\mathcal{S}_{M-L}$, this procedure defines the partition $\pi^*=\{\mathcal{S}^*, \mathcal{S}_1,\ldots\mathcal{S}_{M-L}\}$. Observe that no agent in $\mathcal{S}^*$ can profit from joining a coalitional deviation. Without the help of agents in $\mathcal{S}^*$, no agent in $\mathcal{S}_1$
could profit from joining a coalitional deviation. In general, without the help of agents from earlier groups in the sequence, a profitable coalitional deviation is not possible. Hence there is no profitable coalitional deviation and $\pi^*$ is in the core.

\end{document}